# Optimisation of Quantum Trajectories Driven by Strong-Field Waveforms

S. Haessler[1], T. Balčiūnas[1], G. Fan[1], T. Witting[2], R. Squibb[2], L. Chipperfield[2,3], A. Zaïr[2], G. Andriukaitis[1], A. Pugžlys[1], J. W. G. Tisch[2], J. P. Marangos[2], A. Baltuška[1]

[1] Photonics Institute, Vienna University of Technology, Gusshausstrasse 27/387, 1040 Vienna, Austria
[2] Blackett Laboratory, Imperial College, London SW7 2AZ, United Kingdom
[3] Max Born Institute, Max-Born-Straße 2 A, 12489 Berlin, Germany

**Quasi-free field-driven electron trajectories are a key element of strong-field dynamics. Upon recollision with the parent ion, the energy transferred from the field to the electron may be released as attosecond duration XUV emission[1,2] in the process of high harmonic generation (HHG). The conventional sinusoidal driver fields set limitations on the maximum value of this energy transfer, and it has been predicted that this limit can be significantly exceeded by an appropriately ramped-up cycle-shape[3]. Here, we present an experimental realization of such cycle-shaped waveforms and demonstrate control of the HHG process on the single-atom quantum level via attosecond steering of the electron trajectories. With our optimized optical cycles, we boost the field-ionization launching the electron trajectories, increase the subsequent field-to-electron energy transfer, and reduce the trajectory duration. We demonstrate, in realistic experimental conditions, two orders of magnitude enhancement of the generated XUV flux together with an increased spectral cutoff. This application, which is only one example of what can be achieved with cycle-shaped high-field light-waves, has far-reaching implications for attosecond spectroscopy and molecular self-probing.**

Controlling the shape of the driving laser waveform allows direct steering of the quasi-free electron trajectories at the heart of strong-field phenomena. Technically, this requires the generation of powerful phase-locked more-than-octave spanning spectra, which for the last 20 years had only been possible by combining a fundamental with one of its optical harmonics, typically the second. Early in the development of HHG light sources it was shown that such two-colour fields can enhance the spectral content and brightness[4] of the generated XUV emission. The appearance of even as well as odd order harmonics reflects the additional periodicities in the electric field waveform when a fundamental field and its second harmonic are employed. For the case of a weak second harmonic the additional field interaction has been found to act as a phase gate in HHG that can be used to extract the return times of the recolliding trajectories[5]. Stronger fields have been used in optical gating schemes for isolated attosecond pulse generation[6]. Several factors have been shown to contribute to enhanced brightness in the HHG spectrum: enhanced tunnel ionisation at the "right" instants when recolliding trajectories are launched[7,8], the alteration of the phases[9] of the quantum trajectories responsible for a given frequency high harmonic field, or their coalescence in a spectral caustic[10]. Many examples of this have been studied, including both combinations of a fundamental with an optical harmonic, where relative phase control is possible, and schemes involving incommensurate frequencies with no possibility for systematic phase control[11,12]. The most substantial efficiency enhancements have been reported for orthogonally polarized two-colour combinations of a fundamental with its second harmonic[7,13]. Since in such fields, conditions for the recollision of trajectories have to be satisfied in two independent dimensions, the selection of ionization instants that lead to efficient recollision and thus harmonic emission is more severe than for parallel polarizations. This limits the spectral range over which enhancement of HHG can be achieved; e.g. in ref. 7 to the lowest-order plateau harmonics. Use of orthogonally polarised fields also provides a control over the angle and displacement of the returning electron trajectories[14] which has been shown to be of utility in HHG self-probing of the source atom or molecule[15], and in selection of the dominant trajectory class (short or long)[16]. To gain the full benefit of both enhanced field ionisation and steering of the electron return for HHG efficiency enhancement it is however preferable to use multi-frequency fields with parallel polarisation.

The work on the "perfect wave for HHG"[3] proves that a temporally confined ionisation event coupled to an electric field ramp to ensure the required rapid return of most trajectories will optimize HHG in terms of highest cut-off and efficiency. Despite the numerous successes of two-colour drive fields, it is clear that with only two colours, the scope for waveform (Fourier-)shaping remains rather limited and the efficient synthesis of a waveform approximating to the "perfect wave" ideal requires more frequency components. The obvious next step is to use three phase-locked fields with control of the two independent phase delays. Two extensions in this direction have recently been reported[17,18]. In one case[18], a combination of an 800-nm fundamental with its second and third harmonic allowed the control of the electron trajectory phases for a particular return energy to improve intra-pulse phase matching and enhance the generation of a single harmonic. However, neither were the two relative field phases independently controlled nor was the enhancement demonstrated over a broad spectral range as required for attosecond pulse synthesis. Three broad band field components[17], generated through self-phase modulation in a hollow-core fiber and then separated, individually manipulated and recombined, were used for the sub-femtosecond confinement of field ionization to a single field crest. While the synthesized field has been used for generation of isolated attosecond pulses through HHG, no enhancement or control of this generation via the delays of the three field components has yet been reported. The scaling to higher pulse energies is hampered by the limitations of hollow-core fibres and would thus require costly separate parametric amplification for each colour component. Note that in that method[17], each field component already has a very large bandwidth so the resulting field waveform is hard to analyse in terms of standard Fourier synthesis.

Here we apply shaped optical cycles, synthesized from three discrete spectral bands covering 1.6 octaves, to the complex optimization task of efficient launch, acceleration and return of the electron trajectories in HHG. This addresses a fundamental bottleneck in HHG driven by a sinusoidal driver of wavelength $\lambda$ and intensity $I$, which determine the highest achievable electron energy at recollision, 3.2 $U_p$, where $U_p \propto I\lambda^2$. The laser intensity that can be used is limited to a narrow range due to the combined effect of the exponentially growing ionization rate and saturation (near unity ionization probability). Increasing wavelength leads to a drop $\propto \lambda^{-6}$ of the conversion efficiency[19], as a consequence of increased trajectory excursion duration, $\propto \lambda$, which results in additional quantum wavepacket spreading and so reduces the recollision amplitude. For some applications, phase-matching in long high-pressure media can compensate for this loss[20]. Here, applying shaped optical cycles that follow the principles laid out by the "perfect wave for HHG", as illustrated in Fig. 1, we tailor the single-atom quantum dipole to demonstrate the efficacy of the "perfect wave" concept and so take an important step towards the removal of the above-mentioned bottleneck.

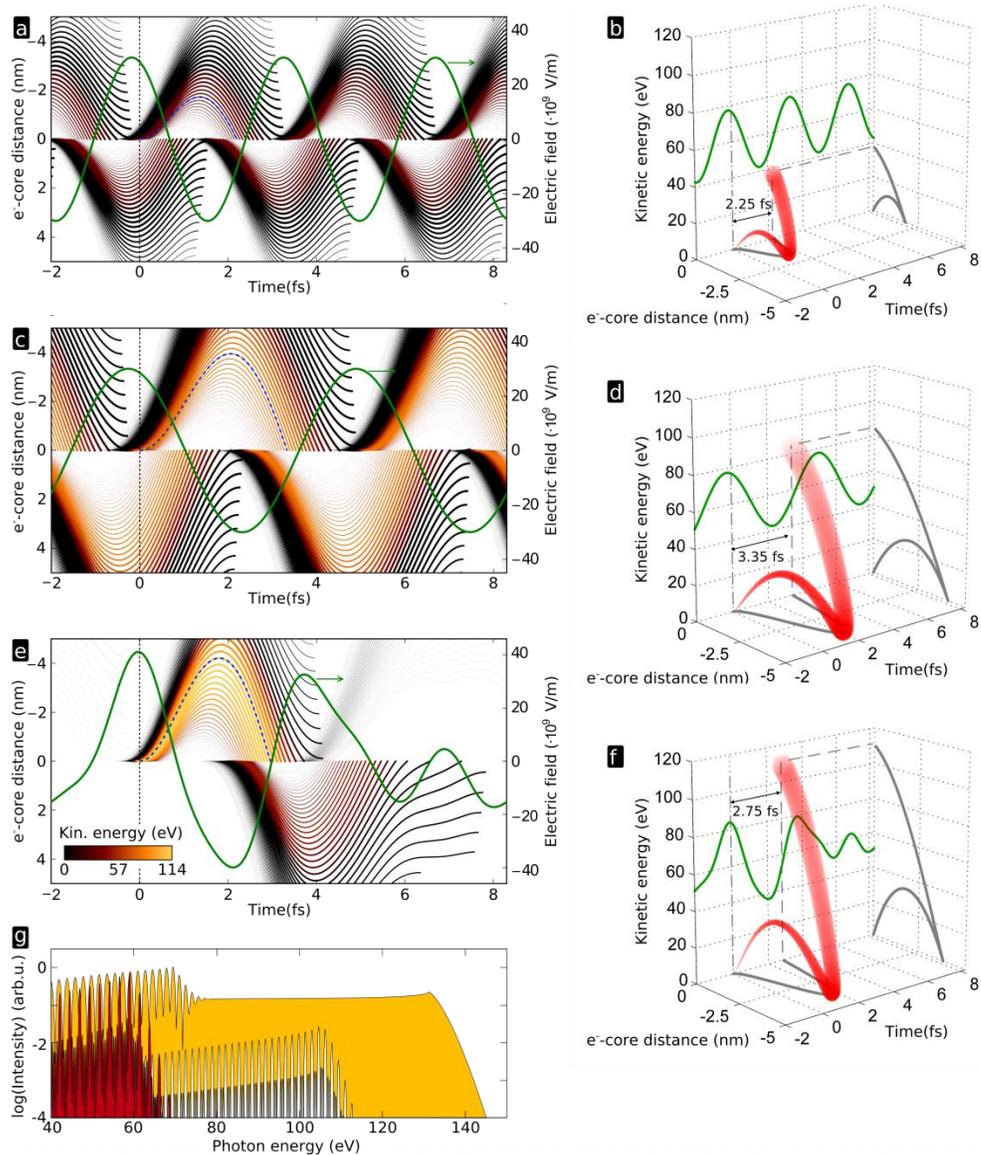

**Figure 1**: Simulated optimization of HHG via the driving waveform. (a)-(f) Classical electron trajectories, with driving electric field shown by the green line, (g) HHG spectra. Both single-colour drivers [1.03 μm (a,b) and 1.545 μm (c,d)] have an intensity of $1.2 \times 10^{14}$ W cm$^{-2}$. The 3-colour waveform [(e,f), combination of 1.03 μm, 0.515 μm and 1.545 μm with relative intensities as in our experiments (Fig. 4c) and optimal phase delays, see also SM4[21]] has the same fluence within its 10.3-fs period as the single-colour waves. In (a),(c),(e), the classical electron trajectories are coloured according to their recollision energy (the colour-bar in (c) is also valid for (a) and (b), non-recolliding trajectories are black). A thicker line indicates a higher tunnelling rate at the ionization instant (normalized in each panel). The cutoff trajectory with the highest recollision energy is marked by the dashed blue line and its ionization instant, t=0, is marked by the vertical dotted line. This cutoff trajectory is again shown in (b),(d),(f) with kinetic energy in the third dimension by points with radii that linearly increase with excursion time, symbolizing wavepacket spreading. In the single-colour drivers, a majority of the continuum electron wavepacket does not return to the core and the cutoff trajectory is rather "weak" (a,c). The longer driver wavelength in (b,c) leads to higher recollision energies but also to longer excursion durations and thus reduced recollision amplitude. In the 3-colour waveform (e,f), the highest recollision energies are achieved, the corresponding trajectories are launched with the highest tunnelling rate and the cutoff excursion duration (2.75 fs) is shortened. Non-recolliding trajectories that do not contribute to HHG but only to detrimental ionization are much weaker. These are the principles of HHG enhancement laid out by the "perfect wave" for HHG[3]. Panel (d) shows the corresponding single-atom HHG spectra (red for 1.03 μm, white for 1.545 μm, yellow for the 3-colour driver) calculated for the ionization potential of argon (15.76 eV) over a 10.3-fs time window and including the short-trajectory class only[21].

We coherently combine three discrete colour components: $\lambda_1$ = 1030 nm from a CEP-locked Yb-based femtosecond laser amplifier, its second harmonic $\lambda_2$ = 515 nm, and $\lambda_3$ = 1545 nm, the signal wave from a white-light-seeded OPA, pumped by the 1030 nm laser (see Supplementary material[21] for details). In the well-proven combination of the optical harmonic with the fundamental, the relative phase delay remains locked independently of the CEP, $\phi_{CEP}$, of the fundamental. This is because in the generation of the *n*th optical harmonic, the complete phase factor of the fundamental is multiplied by *n*, so that a CEP-fluctuation shifts both the fundamental and its harmonic wave by the same phase delay $\phi_{CEP} \lambda_1 (2\pi c)^{-1} = n\phi_{CEP} (\lambda_1/n) (2\pi c)^{-1}$. Adding more optical harmonics becomes increasingly difficult as the conversion efficiency rapidly plunges with order. Furthermore, the frequency up-conversion rapidly reaches into the VUV spectral region, which would create serious practical complications such as absorption in air and multi-photon absorption in the target medium. In contrast, frequency-down-converted colour components, created by OPA, can be generated with high conversion efficiency and reach towards the mid-IR spectral region which is highly advantageous for strong-field interactions. In white-light-seeded OPA, the signal (wavelength $\lambda_s = \lambda_1/m$, $m<1$) "inherits" the CEP from the pump beam ($\lambda_1$), while the idler ($\lambda_i = \lambda_1/(1-m)$) has a passively stable CEP[22]. Thus, when the pump-laser CEP fluctuates, the idler's phase delay does not change, while the pump and signal shift by different phase delays: $\phi_{CEP} \lambda_1/2\pi c \neq \phi_{CEP} \lambda_1/m2\pi c$. In our case, the active CEP-locking of the Yb-based pump-laser ensures stable and controllable relative phase delays of all colour components. While this principle has been recognized before[23], we report here its first realization at sufficient power for driving HHG. While in this proof-of-concept work, we do not yet use the phase-locked 3090-nm idler which is also generated in our OPA, the extendability of our scheme by this additional colour component, as well as by optical harmonics of the signal wave or additional OPA with different down-conversion factor *m* is straightforward. The same is valid for the up-scaling of pulse energies and average power.

Superposing our three colour components, $\lambda_1$, $\lambda_2 = 0.5\,\lambda_1$, and $\lambda_3=(3/2)\lambda_1$, with controlled phase delays, $\tau_2 = \phi_2\,\lambda_2/2\pi c$ and $\tau_3 = \phi_{CEP}(\lambda_3 - \lambda_1)/2\pi c$ of the 515-nm and 1545-nm components relative to the 1030-nm fundamental, respectively, we realize a Fourier synthesis of optical cycles, recurring with a 10.3-fs period under the envelope of a ≈ 180-fs laser pulse with > 0.5 mJ total energy. The main source of rapid phase jitter affects $\tau_3$ and is due to the 0.95-rad jitter of the pump laser CEP. Other factors that lead to averaging over both $\tau_2$ and $\tau_3$ in the macroscopic HHG target are discussed in the supplement[21]. We use our synthesized optical cycles to drive HHG in argon, the result of which is then analysed in a flat-field XUV spectrometer.

In order to study how the optical cycle shape governs the HHG emission, we set the peak intensities of the three colour components at the HHG medium to $3\times10^{13}$ W cm$^{-2}$ (1030 nm), $0.25\times10^{13}$ W cm$^{-2}$ (515 nm), and $1.8\times10^{13}$ W cm$^{-2}$ (1545 nm), and measure the HHG spectra while scanning the phase delays $\tau_2(\phi_2)$ and $\tau_3(\phi_{CEP})$. To reveal the impact of the optical shaping on the trajectories of the recolliding continuum electron wavepacket (REWP), we have normalized the measured HHG spectra by the squared recombination dipole matrix element for argon[21,24,25]. Furthermore, in order to highlight the fast sub-cycle dynamics, we have filtered out the dense 0.4-eV harmonic peak modulation[21] corresponding to the 10.3-fs period of our waveforms. The result can be considered as an REWP spectrum generated within a single optical cycle of the driving wave.

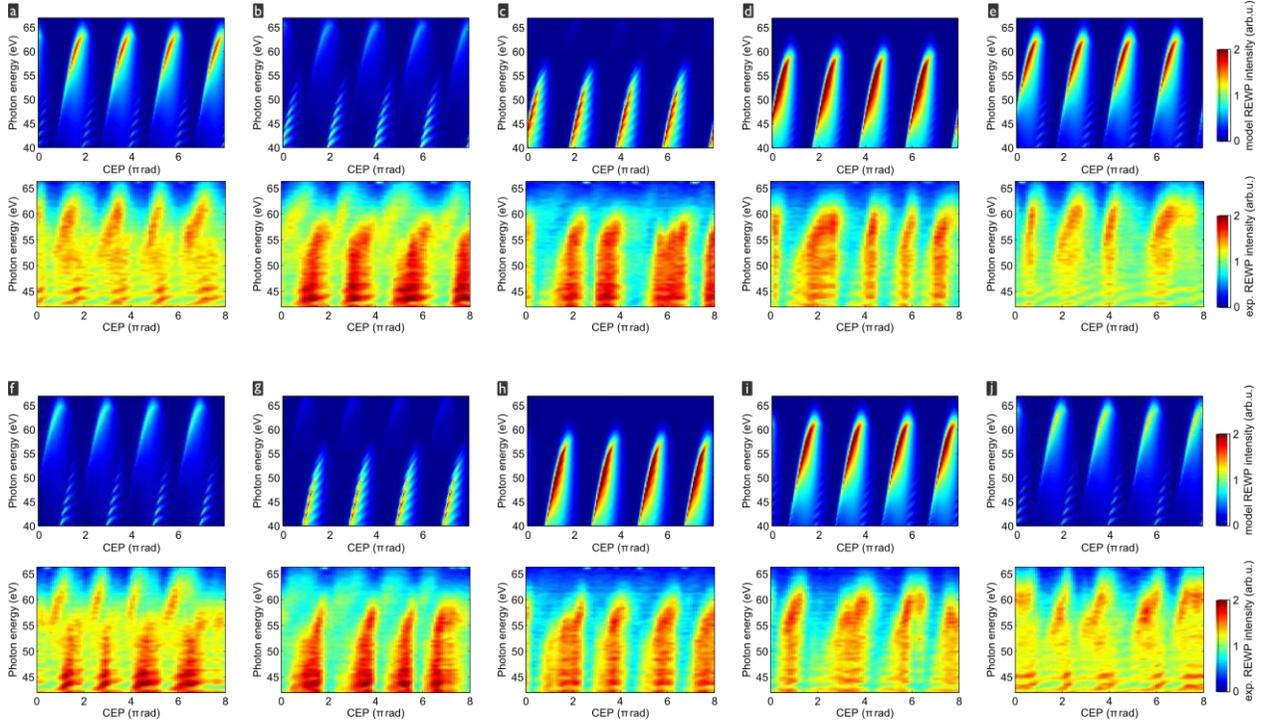

**Figure 2**: (a) - (j) Simulated (upper) and measured (lower) REWP spectrograms for $\phi_2$ = 1.19π, 1.42π, 1.65π, 1.88π, 0.11π, 0.34π, 0.57π, 0.80π, 1.03π and 1.26π. The simulations have been performed for a single 10.3-fs long optical cycle using the same intensities of the three colour components as estimated for the measurements. Only the short trajectory family has been included[21], since this will be best phase matched in a macroscopic medium. The CEP-values in the experiment have an unknown offset, which could slip between scans for different $\phi_2$. Note that $\phi_2$ may be considered as jitter-free whereas $\phi_{CEP}$ jitters by ≥0.95 rad r.m.s.[21]. HHG spectra have been measured for π/10 steps of $\phi_{CEP}$ averaging over 100 shots. The acquisition time of one spectrogram was thus below 10 s.

Figure 2 shows these data as function of the laser CEP, $\phi_{CEP}$, for ten values of $\phi_2$ covering a 2π range, together with REWP-intensities simulated with the Lewenstein model[26] using the quantum path analysis[21,27]. These simulations are performed for a single atom and a single 10.3-fs long optical cycle and do not aim at reproducing in detail the experimental data (see SM6 for further discussion), but at providing a reference for qualitative structures in the spectrograms (photon energy vs. $\phi_{CEP}$) as signatures of trajectory steering at the sub-cycle, single-atom level. We find good agreement of the overall structures with those observed experimentally for all $\phi_2$ values. The spectral modulation appearing for some values of $\phi_2$ at photon energies below ≈ 55 eV has the same ≈ 2-eV period in the experimental and simulated spectrograms (see Figs. S4 and S6). Our calculated quantum trajectories show that it corresponds to the interference of two recollision events per optical period, separated by ≈ 2 fs (see, e.g. the two events in Fig. 3a and b). There is no such modulation at the highest photon energies since these are produced only in a single recollision per optical cycle. The observed agreement is strong evidence that, despite experimental phase delay jitter[21], we control the HHG process on the single-atom level by directly steering electron quantum trajectories with our shaped optical cycles on the attosecond time scale.

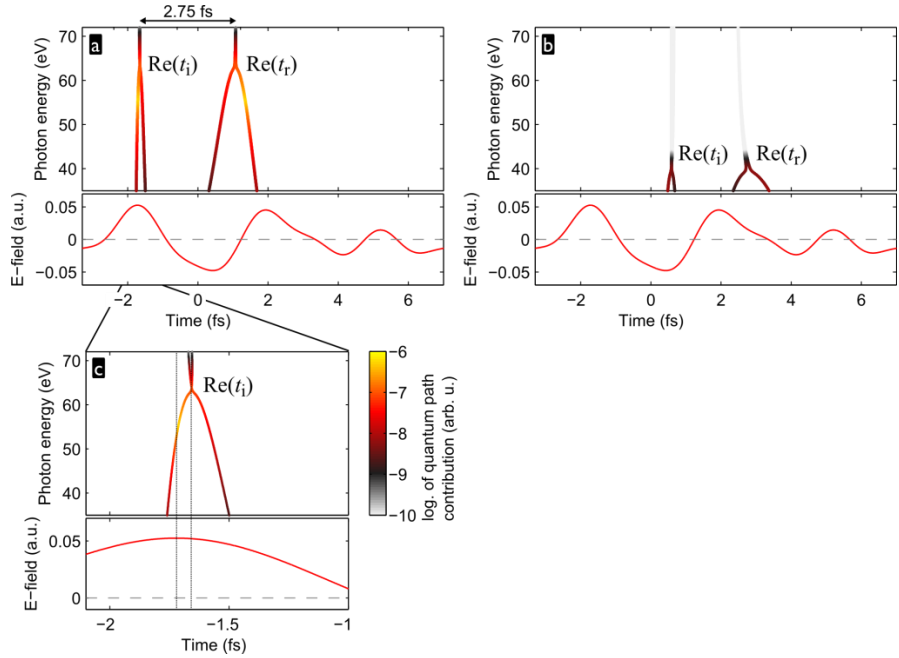

**Figure 3**: Results of the saddle point analysis of the Lewenstein-model simulations for the parameters of Fig.2 with $\phi_2$=1.19π and $\phi_{CEP}$=1.9π, shown together with the driving waveform on the same time axis. In panels (a) and (b), a full optical cycle of the driving waveform is covered. Two recollision events corresponding to HHG emission above 35 eV occur, shown separately in panels (a) and (b). Each event is represented by a matching pair of arches, the left and right of which represent the real parts of the ionization and recollision instants of the quantum trajectories, respectively. The left and right branches of the arches of ionization instants, joining at the spectral cutoff, correspond to the long and short trajectory class; the same applies vice versa to the arches of recollision instants (see also Fig. S2). The data points are colour-coded, showing in logarithmic scale the contribution of the corresponding quantum path to the emitted HHG intensity. Panel (c) shows a zoom onto the ionization instants. The ionizing field strength launching the cutoff trajectories (at right vertical dashed line) is still 99 % of its peak value (at left vertical dashed line). While the long trajectories have higher contribution in this single-atom calculation, phase matching in a macroscopic medium usually leads to a clear dominance of the short-trajectory contribution.

The composite waveforms used in this work have the potential to significantly enhance HHG. As visible in Fig. 2, the waveform corresponding to $\phi_2$ = 1.19π and $\phi_{CEP}$ = 1.9π leads to good efficiency at high photon energies. Figure 3 shows the ionization and recollision instants of the quantum trajectories calculated for this waveform as well as the contribution of each trajectory to the HHG emission[21]. Only a single recollision event leads to the emission of photons above 45 eV. This is because the available driving field energy is concentrated to produce essentially two strong field crests per 10.3-fs optical period which can efficiently drive HHG. In a single-colour wave with the same fluence per 10.3-fs time window, the attainable peak field strengths would always be lower, leading to reduced efficiency. In the synthesized waveform, the high-energy trajectories are launched by the strongest field crest, and as shown in Fig. 3c) the field crest maximum is very efficiently used for launching the recolliding trajectories for HHG. In contrast to sinusoidal driving fields where all recolliding trajectories are launched *after* the field crest peak, the asymmetry between the ionizing and subsequent accelerating field crest in the synthesized waveform shifts the ionization instants to *around* the field crest maximum (similar to the ionization enhancement in Ref. 28). The waveform after the ionization instants leads to an improved energy transfer from the driving field to the accelerated electron and thus enhances the recollision energies and thus the attainable HHG photon energy cutoff. The synthesized waveform leads to relatively short excursion durations of < 2.75 fs for the relevant trajectories (for a single-color driver of, e.g., 1545-nm wavelength, the excursion duration for the cutoff trajectory would be significantly longer (3.35 fs), without however leading to higher recollision energies). This shape, a strong and sharp ionizing field crest followed by a ramped-up longer field crest is very similar to that described for the "perfect wave" for HHG[3]. Our quantum-path analysis thus confirms the enhancement principles already highlighted by the classical trajectories shown in Fig.1c), which remain in effect even when the relative fluencies of the colour components are not optimal.

The experimental HHG data shown in Fig. 2 have been obtained under conditions of fairly low ionization probabilities far from saturation. To demonstrate the full benefit of our shaped optical cycles for the cutoff and efficiency enhancement of HHG, we have repeated our experiments at higher driving intensity and thus in conditions of high practical relevance close to saturation, which implies a reduced XUV phase matching bandwidth and potentially some reshaping of the driving waveforms during propagation in the HHG medium. We have also increased the relative strength of the 1545-nm component to further approach the optimal waveform theoretically predicted for our three colour components (see SM4[21]). The peak intensities of the three colour components in the HHG medium have thus been set to approximately $5.7\times10^{13}$ W cm$^{-2}$ (1030 nm), $0.3\times10^{13}$ W cm$^{-2}$ (515 nm), and $6\times10^{13}$ W cm$^{-2}$ (1545 nm). We compare the HHG spectra obtained with our shaped optical cycles to HHG with single-colour drivers of 1545 nm and 1030 nm wavelength.

For this experimental study, summarized in Fig. 4, we do not compare the different driver waves at constant fluence during a given period (as in the simulations of Fig.1), but rather at their respective practical limits in our experimental set-up which set the experimentally relevant benchmarks. All experimental conditions except for the pulse energies were left exactly the same for the single-colour benchmarks as for the synthesized waveforms. The high-energy 1030-nm pump laser easily drives HHG into saturation[21], which thus sets the practical limit. HHG saturation by the OPA output at 1545 nm was not achieved due to energy restrictions and wavelength scaling. We have optimized the OPA for highest conversion efficiency and output beam quality, so using the full OPA output sets the practical limit for HHG. For the synthesized waveforms, we have performed a similar scan[21] as shown in Fig.2 and then chose the waveform ($\phi_2=1.8\pi$ and $\phi_{CEP}=0.3\pi$) that produced the highest HHG signal between 60 eV and 72.5 eV (Al-filter edge). This is not the same optimal waveform as shown in Fig.1, which would maximize the HHG efficiency above 100 eV and thus beyond the experimentally detected spectral range.

Fig.4 shows strong enhancement of HHG with the synthesized optical cycles. Whilst with the 1545-nm short-wavelength-IR (SWIR) driver, a fairly high spectral cutoff is achieved, the HHG flux is very low. On the other hand, in HHG driven to the onset of saturation by the 1030-nm pump laser, we cannot shift the cutoff beyond 55 eV[21]. Figs.4a,b thus clearly demonstrate the limitations of sinusoidal driver waves. In comparison, the synthesized optimal waveform generates an HHG spectrum that unites high spectral intensities with a cutoff well beyond the saturation limit of the efficient near-IR driver. In terms of XUV spectral intensity, the optimal waveform is on par with the near-IR driver and leads to a >80 enhancement factor compared to the SWIR driver. The denser harmonic comb spacing leads to even greater enhancement in the integral XUV flux (measured factor >140 in the 55-65-eV range). According to our simulations (Fig. S10 [21]) for the synthesized waveform, the dominant recollision events occur once per 10.3 fs (it would take a 6.2-μm single-colour driver to realize the same periodicity), as opposed to once per 2.6 fs in the SWIR case. Consequently, we would expect a several hundred times enhancement in the flux per attosecond burst and thus great implications to future sources of high-energy (isolated) attosecond pulses.

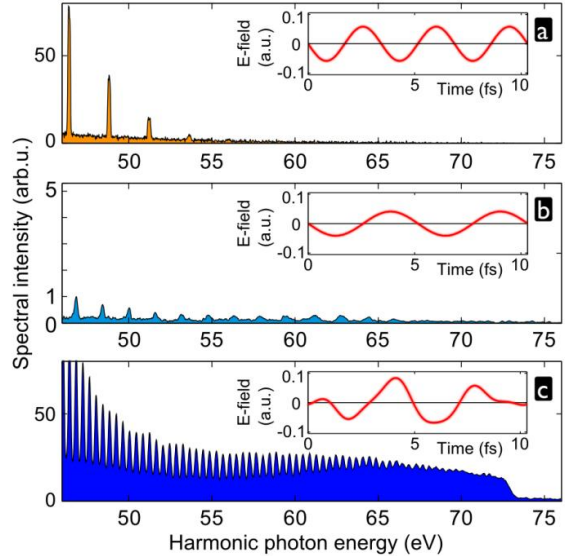

**Figure 4**: Experimental HHG spectra generated in argon with different driver waves, shown in the insets. The spectral intensities are given on the same scale and are thus directly comparable. (a) 1030-nm driver with the same 0.54 mJ pulse energy as the three-color pulses, leading to the onset of saturation in HHG[8]. (b) 1545-nm driver with the full OPA output pulse energy (0.35 mJ). (c) Chosen 3-colour driver ($\phi_2$=1.8π and $\phi_{CEP}$=0.3π) giving the highest HHG signal between 60 eV and the Al-filter absorption edge.

We conclude that under realistic experimental conditions, the capability of quantum trajectory engineering via the shaping of the driving optical cycles provides a powerful tool to significantly enhance HHG. By generating a waveform that implements the principles of the theoretically proposed "perfect wave for HHG"[3], we avoid the cutoff-vs.-efficiency dilemma[19] of HHG with sinusoidal driver waves.

The HHG driving performance of our waveforms goes well beyond what has been reached before. In the two-colour method of ref. 7, the aggravated selectivity for the return of launched electron trajectories required a driving intensity of ~$10^{15}$ W cm$^{-2}$ to demonstrate enhancement of 2 orders of magnitude at photon energies up to 72 eV. Even higher intensities will barely be applicable with neutral gases. With our method, we already demonstrate the same enhancement of spectral intensity in the same spectral region in a pure proof-of-concept study at one order of magnitude lower driving intensity. This advantage is achieved by "targeted" trajectory steering in our favour, following the guidelines of the "perfect wave"[3] for HHG. Furthermore, we can with good reason infer from our simulations that similar enhancement is possible at higher photon energies (Fig. 1) than those detected in our experiment (macroscopic phase matching provided). The enhancement will further improve when phase delay jitter is reduced. There is no principle obstacle in scaling our method towards much higher photon energies by increasing the intensities or adding the mid-IR idler wave, albeit technical challenges like phase-matching at high photon energies and possibly shortening of the pulse envelopes[29] will have to be met. Finally, our simulations let us infer that our method has the important advantages of greatly increasing the periodicity of the attosecond pulse train (by a factor 4, as compared to factor 2 in the case of ref. 7) and of decreasing the attochirp (through the reduction of the cutoff excursion times, whereas these actually increase in the case of ref. 7).

Our demonstrated optimization holds significant potential for applications of HHG. Without sacrificing signal intensity for the self-probing of atoms and molecules[30], the desired higher bandwidths of the REWP can be attained even at moderate laser intensities, which is an important requirement for the study of larger molecules or clusters. The very dense harmonic spacing can be useful for higher spectral resolution in, e.g., determining the position of Cooper minima[24], dynamic or structural minima in HHG from aligned molecules[31], features due to resonances and in applications to high resolution transient X-ray absorption measurements. The corresponding large temporal spacing of the attosecond pulses is of interest in gating techniques for selecting isolated attosecond pulses[6]. The reduced atto-chirp is highly advantageous for the generation of ever shorter and more intense attosecond pulses. It can be compensated with thinner, and thus less absorbing metal filters[32], thereby further increasing the attainable attosecond pulse energy on target.

Our HHG enhancement results demonstrate a new example of the many possible applications of advanced shaping of optical cycles. Generally, any directly laser-field driven process can be optimized by adequate intra-cycle pulse shaping. We expect new possibilities to emerge in a broad range of laser-matter interaction regimes, involving, e.g., Brunel electrons whose field-driven trajectories can lead to THz-emission[33], plasma heating and HHG on plasma mirrors[34], or particle acceleration[35].

**Acknowledgements**

These studies were supported by the ERC (projects CyFi 280202 and ASTEX 290467) and the EPSRC (grants Nos. EP/I032517/1, EP/E028063/1 and EP/F034601/1). A. Zaïr acknowledges support from EPSRC grant EP/J 002348/1. S. Haessler acknowledges support by the EU-FP7-IEF MUSCULAR and the by the Austrian Science fund (FWF) through grant M1260-N16.

**Author Contributions**

S.H. and T.B. contributed equally to this work. L.C. developed the underlying original "perfect wave" theory. A.B., J.P.M., J.W.G.T. and L.C. conceived the concept for this work. A.B., T.B., G.A., A.P., G.F. and S.H. designed and developed the waveform synthesizer. S.H., T.B., A.B and R.S. designed and developed the HHG set-up. S.H., T.B., G.F., T.W. and A.Z. performed the HHG experiments and the data analysis. S.H. made the quantum path simulations. L.C. made additional supporting HHG simulations and optimization calculations. S.H., J.P.M., T.B., G.F. and A.B. wrote the manuscript. All authors discussed the results and the manuscript.

# Supplementary material to
## "Optimisation of Quantum Trajectories Driven by Strong-Field Waveforms"

S. Haessler[1], T. Balčiūnas[1], G. Fan[1], T. Witting[2], R. Squibb[2], L. Chipperfield[2,3], A. Zaïr[2], G. Andriukaitis[1], A. Pugžlys[1], J. W. G. Tisch[2], J. P. Marangos[2], A. Baltuška[1]

[1] Photonics Institute, Vienna University of Technology, Gusshausstrasse 27/387, 1040 Vienna, Austria
[2] Blackett Laboratory, Imperial College, London SW7 2AZ, United Kingdom
[3] Max Born Institute, Max-Born-Straße 2 A, 12489 Berlin, Germany

## 1 Experimental setup

The fundamental technological challenge for shaping the optical cycle of the carrier wave is the creation of a phase-locked spectrum spanning more than an octave. However, for the shaping of a periodically repeating optical cycle under a "longer" femtosecond envelope, the more-than-octave spanning spectrum need not be continuous but may consist of discrete colour components.

We create the base colour components for our waveform synthesizer in a collinear optical parametric amplifier (OPA). The full phase locking (i.e. shot-by-shot stable phase delays between the colour components), which is obviously a crucial prerequisite for a Fourier-synthesis of waveforms, is achieved as follows: in white-light-seeded OPA, the signal (frequency $\omega_s$) "inherits" the CEP, $\phi_{CEP}$, from the pump beam ($\omega_p$), while the idler ($\omega_i$) has a passively stable CEP $\phi_i$ =const.[1]. Thus, when the pump-laser CEP fluctuates, the idler's phase delay does not change, while the pump and signal waves shift by different phase delays: $\phi_{CEP}/\omega_p \neq \phi_{CEP}/\omega_s$. The three waves, pump, signal and idler, that are the building elements of our Fourier-synthesized waveform are thus slipping with respect to each other and the waveform will change with each laser shot; unless of course the pump laser CEP, $\phi_{CEP}$, is locked.

The pump laser technology based on Yb:CaF$_2$ provides the required energy level for driving multicolor HHG. The general scheme of the laser amplifier system is depicted in
Figure **S1**(a). An Yb:KGW CPA laser system ("Pharos" by Light Conversion Ltd.) is used as a front-end for a high energy cryogenically cooled Yb:CaF$_2$ regenerative amplifier[2] (RA) [1]. The front-end CPA amplifier delivers pulses stretched to 200ps at up to 6 W of average power at repetition rates of 10-200 kHz and up to 1 mJ pulse energy at 1 kHz. For the purpose of seeding the second-stage amplifier, the energy after the front-end was kept at a lower, 100-μJ level. The output pulse energy after the Yb:CaF$_2$ RA stage is boosted up to 6 mJ because of the advantage of a much lower third order nonlinearity of Yb:CaF$_2$ as compared to other Yb host materials which are less suitable for energy scaling owing to of nonlinear effects during the amplification. Using an electronic phase-lock loop (PLL), the carrier-envelope-offset phase of the amplifier chain is actively stabilized. The resulting r.m.s. phase jitter from the first stage amplifier is[3] 0.45 rad and the typical jitter of the CEP after the second stage RA is approximately 0.95 rad. The central wavelength of the amplified laser output is $\lambda_p$ = 1.030 μm and the pulse duration has been measured with second harmonic generation frequency-resolved optical gating (SHG FROG) to be 180 fs.

The combination of white-light-seeded OPA and multi-colour interferometer which makes up our waveform synthesizer is schematically shown in Fig S1(b). The white-light seed is generated in a 1 mm thick sapphire plate. The OPA consists of three stages with type-II phase matching in KTP crystals and is tuned to amplify a signal wave at $\lambda_s$ = 1.545 μm, such that the idler wave is at $\lambda_i$ = 3.09 μm (chosen to be the longest possible idler wavelength for which KTP is still transparent). An interference-filter after the first OPA stage narrows the bandwidth around 1545-nm so that the final amplified signal pulse is nearly as long (170-fs) as the pump pulses and, most importantly, is close to Fourier-limited. In our current waveform synthesis work, the long-wavelength idler wave is not used.

Using dichroic optics, the signal wave is separated from the OPA output and then combined with a "fresh" pump laser pulse obtained by splitting off part of the laser output before the OPA. The phase delay $\tau_3$ (as defined in the main text) of the signal wave relative to the pump wave is scanned by varying the locking point of the CEP of the pump laser amplifier slow-loop.

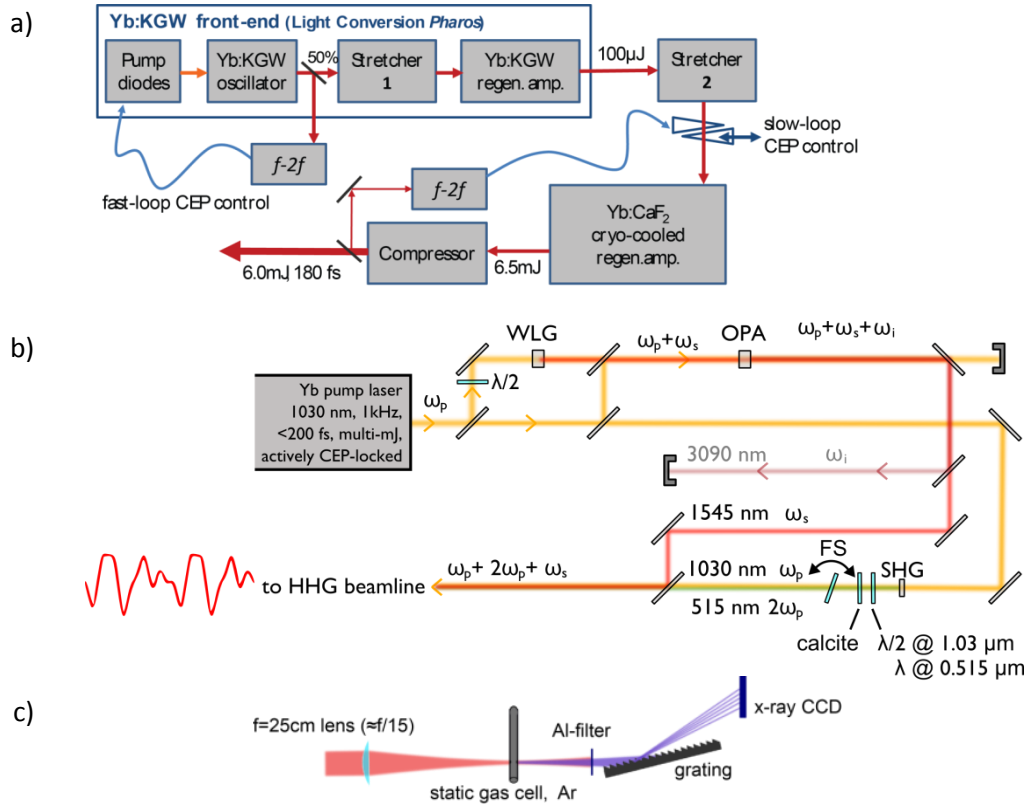

**Figure S1**: Implementation of our waveform synthesizer for HHG driving. a) a multi-mJ diode-pumped femtosecond 1-μm CEP-stabilized pump laser, b) collinear mid-IR CEP-stable OPA followed by multi-colour interferometer with sub-cycle stability, c) schematic of the HHG beam-line.

As a third colour component, the second harmonic of the pump laser pulse at $0.5\lambda_p$ = 0.515 μm is generated by frequency doubling a small part of the pump beam using a thin (0.5 mm) BBO crystal. The polarizations of the second harmonic and the fundamental are matched using a multi-colour waveplate (λ for 0.515 μm, λ/2 for 1.03 μm). This also matches the polarizations to that of the signal wave. The group velocity delay between the pump wave and its second harmonic is compensated by inserting a calcite plate of ≈ 1.5 mm thickness. The fine tuning of the phase delay $\tau_2$ (as defined in the main text) between the fundamental wave and its second harmonic is done by rotating a thin (1-mm) fused silica plate.

The multi-colour field, synthesized after the interferometer, thus consists of three pulses with parallel polarizations and carrier wavelengths of 1.545 μm, 1.03 μm and 0.515 μm. The three spatially and temporally superposed pulses are then focused by an f=25 cm lens (≈f/15) about one Rayleigh length (≈ 1 mm) before a nickel tube with 0.8 mm inner diameter, which is pierced by the laser beam. The tube is backed with ≈ 200 mbar of argon. A 100-nm Al-filter downstream transmits XUV radiation between 15.5 eV and 72.5 eV. This high-order harmonic emission is measured using a flat-field XUV spectrometer, consisting of an entrance slit, a Hitachi grating No. 001-0640 and an X-ray CCD camera (Andor Newton 920 SO) as depicted in Fig. S1(c).

We determine the intensities of the individual colour components on target by measuring both their their focal spot diameters directly on a CCD camera and their pulse energies. With the known pulse durations of the pump and signal pulses, as well as the assumption that the second harmonic pulses are about 170 fs long, we can estimate the intensities at focus or some distance behind. While this gives reliable estimates for the *relative* intensities of the colour components, the not precisely known distance of the HHG target from the actual beam waists is the main source of uncertainty for the absolute values. These are then more finely set by matching HHG simulations to measured spectra.

## 2 Waveform stability

In this section, we discuss several factors affecting the stability of our generated waveform. These factors each depend sensitively on the experimental conditions and so are hard to quantify precisely. We can however say that in the conditions of Fig.2 of the main paper, the main source of instability is the CEP jitter and concerns the phase delay, $\tau_3$, between the 1.03-µm and 1.545-µm components. In the higher-intensity conditions of Fig. 4, dispersion in the ionizing HHG medium will lead to an additional non-negligible averaging over both relative phase delays $\tau_2$ and $\tau_3$. Nonetheless, the fact that we do observe clear modulations of the HHG signal with the controlled phase delays of the colour components (see Figure S9) is proof of sufficiently stable waveforms.

### 2.1 Shot to shot

From shot to shot, it is mainly the jitter of the two phase delays, $\tau_2(\phi_2)$ and $\tau_3(\phi_{CEP})$ that affects the waveform stability. Thanks to the in-line generation and propagation of the second harmonic, the phase delay $\tau_2$ can be considered as jitter-free. The rather larger pump laser CEP jitter of 0.95 rad r.m.s is the main source of instability; nevertheless this is a jitter of the phase delay $\tau_3$ of only 260 as. Intensity-to-phase coupling in the WLG step of the OPA setup in principle may add to this jitter, but since we have found the pulse-energy fluctuations of the Yb:CaF$_2$ RA to be ≤ 1%, we do not expect a large contribution from this. Finally, the arm-length stability of the multi-colour interferometer (with 1.03-µm in one arm and the 1.545 µm in the other, see Figure S1) will of course affect $\tau_3$. By means of non-linear interferometry of the combined signal and pump pulses, we have determined an upper bound of 150 as r.m.s. for the $\tau_3$-jitter up to that imposed by the common CEP. We expect this to be dominated by the free-running interferometer arm-length jitter. This variation is rather slow (~1 Hz), i.e. not on the shot-to-shot level. Its effect on the rapidly measured spectrograms (e.g. Fig. 2 of the main paper) is thus not so much an averaging along the horizontal dimension but rather deviations from the linearity of the imposed phase-delay ramp. An active stabilisation will certainly be possible.

### 2.2 Under the spatio-temporal pulse envelope

Under ideal conditions, the shaped optical cycle repeats under the femtosecond envelope with a period of 10.3 fs. If the temporal phases of the colour components are not exactly parallel (e.g. because they are not perfectly Fourier-transform-limited), the phase delays $\tau_2$ and $\tau_3$ and thus the optical cycle shape will vary under the femtosecond pulse envelope. Our FROG measurements of the 1.03-µm and 1.545-µm pulses however assure us that such deviations are small within the pulses' FWHM and only become important in the wings, which are unimportant for HHG due to the high nonlinearity of the process.

The same effect occurs for non-parallel wavefronts of the colour components throughout the HHG medium. We optimize the OPA not purely for highest conversion efficiency but also take care of very good spatial quality of the generated signal beam. A telescope in the signal-wave-arm of the interferometer (not shown in Figure S1) collimates the signal beam and sets its diameter to be a factor ≈ 1.2 larger than the pump beam after the interferometer. This leads to approximately equal confocal parameters of the two colour components when they are later focused by the same lens. This means that their relative fluence stays constant and their Gouy phases are equal during propagation, thus keeping a constant phase delay through the beam waists. Also, their wavefront curvature is the same. The same conditions are approximately fulfilled for the second harmonic beam since its beam diameter is smaller than that of the pump beam due to the nonlinear conversion efficiency. Again, the high non-linearity of the HHG process makes distortions in the spatial wings unimportant.

### 2.3 Reshaping during propagation

In the ionizing HHG medium, the different spectral components of our waveforms travel at different phase velocities. Depending on the experimental conditions, this leads to a reshaping of the waveforms. This is mainly due to plasma dispersion while gas gives only negligible dispersion compared to plasma.

The phase delay induced by plasma dispersion for each color component with wavelength $\lambda_i$ is $\tau(\lambda_i) = Lc^{-1} n_{plasma}(\lambda_i) = Lc^{-1} r_e N_e (2\pi)^{-1} \lambda_i^2$, where L is the interaction length, c is the speed of light, $n_{plasma}$ is the refractive index of a free-electron gas, $r_e$ is classical electron radius, and $N_e$ is the free electron density. For a medium with 200-mbar and a 5% ionization fraction, we assume the free electron density to be a constant $N_e = 2.5 \times 10^{17}$ cm$^{-3}$ over the whole 0.8 mm inner diameter of the gas tube. We obtain a phase delay $\Delta\tau_1 = \tau(\lambda_{1\mu m}) - \tau(\lambda_{0.5\mu m})$ between 1.03 μm and 0.515 μm of 240 as, and a delay $\Delta\tau_2 = \tau(\lambda_{1.5\mu m}) - \tau(\lambda_{1\mu m})$ between 1.545 μm and 1.03 μm of 400 as. These values may be more easily interpreted when transformed into fractions of the respective color component's period: $\Delta\varphi_1 = 2\pi\,\Delta\tau_1\,T_{1\mu m}^{-1} = 440$ mrad and $\Delta\varphi_2 = 2\pi\,\Delta\tau_2\,T_{1.5\mu m}^{-1} = 490$ mrad. The propagation-induced phase delay shifts are thus small (< 8%) compared to the colour components' periods—but they are not negligible either. Note that the assumed 5% is already a rather high value for the ionization fraction and best phase matching will require a lower free-electron density. Our estimated phase shifts can thus be considered as "worst case". Even then, while the corresponding waveform reshaping will reduce the contrast of the HHG intensity modulation, we can still consider our synthesized waveforms to remain effectively stable in our experimental conditions (which are typical for HHG experiments with short targets). This is consistent with our observation of clear modulations of the HHG signal with the controlled phase delays of the colour components.

## 3 Quantum path simulations

Here we give a brief description of the theoretical model used in this work for comparison with experimentally measured spectra and assigning the driving waveform to the spectrum. All equations are in atomic units. The intensity, $I_{HHG}$, of the radiation with frequency $\omega$ emitted by an atom driven by a laser field is given by:

$$I_{HHG}(\omega) = \omega^4 |x(\omega)|^2,$$

where $x(\omega)$ is the Fourier-transformed expectation value of the time-dependent atomic dipole moment. Following the fully quantum Lewenstein-model for HHG with the stationary phase (or saddle-point) approximation, developed in refs. 4 and 5, we write:

$$x(\omega) = \sum_s \frac{i2\pi}{[\det(\partial^2(S+\omega t))]^{1/2}} \left[\frac{\pi}{\varepsilon + i(t_r - t_i)/2}\right]^{3/2} \mathbf{E}(t_i)\mathbf{d}[\mathbf{p}_s + \mathbf{A}(t_i)]\,\mathbf{d}^*[\mathbf{p}_s + \mathbf{A}(t_r)] \times \exp[iS(p_s,t_i,t_r) + i\omega t_r], \quad (1)$$

where the triplets ($\mathbf{p}_s$, $t_i$, $t_r$) of stationary canonical momentum, ionization instant and recollision instant define the quantum trajectories of the continuum electron over which the sum in eq. (1) runs. In this equation, $\mathbf{d}[k]$ is the dipole matrix element for bound-free transitions, the star denotes complex conjugation,

$$\mathbf{A}(t) = -\int_{-\infty}^{t} dt''\,\mathbf{E}(t'')$$

is the vector potential of the laser field $\mathbf{E}(t)$, and the quasi-classical action

$$S(p_s, t_i, t_r) = -\int_{t_i}^{t_r} dt' \left[\frac{[\mathbf{p}_s + \mathbf{A}(t')]^2}{2} + I_p\right],$$

with the atomic ionization potential $I_p$ (in our case $I_p$ = 15.76 eV for argon).

The determinant

$$\det[\partial^2(S+\omega t)] = \left(\frac{[\mathbf{p}_s + A(t_r)][\mathbf{p}_s + A(t_i)]}{t_r - t_i}\right)^2 - \left(-\frac{2(\omega - I_p)}{t_r - t_i} + E(t_r)[\mathbf{p}_s + A(t_r)]\right)\left(\frac{2I_p}{t_r - t_i} - E(t_i)[\mathbf{p}_s + A(t_i)]\right),$$

and $\varepsilon$ is a small regularization constant (we set $\varepsilon = 10^{-6}$). The quantum trajectories are found by solving the system of three saddle point equation resulting from setting $d(S + \omega t) = 0$:

$$\frac{-\boldsymbol{\alpha}_{t_i, t_r}}{t_r - t_i} = \mathbf{p}_s, \quad \text{with} \quad \boldsymbol{\alpha}_{t_i, t_r} = \int_{t_i}^{t_r} dt'\,\mathbf{A}(t'), \quad (2)$$

$$[\mathbf{p}_s + \mathbf{A}(t_i)]^2 + 2I_p = 0, \quad (3)$$

$$[\mathbf{p}_s + \mathbf{A}(t_r)]^2 + 2I_p - 2\omega = 0. \quad (4)$$

We solve this system of equations numerically, using classical solutions for ionization and recollision instants as initial guesses. The solutions ($\mathbf{p}_s$, $t_i$, $t_r$) are in general complex valued.

When calculating $I_{HHG}(\omega)$ from these solutions via eq. (1), we have used hydrogenic dipole matrix elements, $\mathbf{d}[k] \sim k /(k^2 + 2I_p)^3$, in Figures 1 and Figure **S3**. All other figures represent the recolliding electron wavepacket, i.e. we neglect the influence of atomic structure (apart from the value of $I_p$) by setting the dipole matrix elements $\mathbf{d}[k]$ to unity. The loss of normalization is unproblematic since we are not interested in absolute values of the calculated dipole. As the ionization-matrix-element $\mathbf{d}[\mathbf{p}_s + \mathbf{A}(t_i)]$ is constant for all saddle-point solutions, due to eq. 3, setting it to unity has no effect on the structure of the spectrum. The structure of the recombination matrix element $\mathbf{d}^*[\mathbf{p}_s + \mathbf{A}(t_r)]$, however, is a major factor governing the shape of the final HHG spectrum. Replacing it by a constant thus leads to an HHG spectrum cleared of this influence. Alternatively, we can interpret the resulting $I_{HHG}(\omega)$ in the spirit of the "quantitative rescattering theory" (QRS)[6] as the intensity of the recolliding continuum electron wavepacket. In the QRS, the full calculated dipole $x(\omega)$ is divided by the recombination dipole matrix element $\mathbf{d}^*(\omega)$, where the link between $\omega$ and $\mathbf{k} = \mathbf{p}_s + \mathbf{A}(t_r)$ is given by eq. (4). Since in eq. (1), the recombination dipole matrix element $\mathbf{d}^*[\mathbf{k}]$ could be factored out of the sum (eq. (4) fixes for a given $\omega$ the same $\mathbf{k} = \mathbf{p}_s + \mathbf{A}(t_r)$ for all quantum paths), the replacement by unity has the same qualitative effect as the division in the QRS.

While computing $I_{HHG}(\omega)$ without the stationary phase approximation is perfectly feasible and in fact a widely used standard method for HHG simulations, we have chosen to take the approach via the quantum trajectories since this allows a straightforward connection to classical electron trajectories and thus better physical understanding of the results. The "quantum path contribution" which is encoded by the coloured data points in Figure 3 of the main paper and Figure S2 and Figure **S10** is equal to the value of $I_{HHG}(\omega)$ calculated for the respective quantum path only. The full $I_{HHG}(\omega)$ results from the interference of quantum paths in the sum (1).

One can easily select certain classes of trajectories from the solutions and thus include only those contributions from the single-atom dipole that can be expected to be well phase-matched in a macroscopic HHG medium. For sinusoidal drivers, the division into short and long trajectories is clear and it is known that under common experimental conditions, the short trajectory class by far dominates the macroscopic signal. In our shaped waveforms, there can be several different recollision events (groups of trajectories starting around the same field crest and recolliding throughout a continuous time window) during an optical cycle. Within each of these events, it turns out that we can still clearly separate long and a short trajectory branches. It is however not obvious if all recollision events should be considered on equal footing. The short trajectory branch of recollision event A might have longer excursion times than the long trajectory branch of event B. The excursion times are not the only factor; a higher field at the ionization instant leads to a "broader tunnel" and thus less initial transverse momentum spread of the launched electron wavepacket and thus less pronounced transverse spread during the excursion. For simplicity, we make the ad-hoc choice for our simulations to simply pick out the short-trajectory branch of each event and otherwise make no additional weighting. Figure S2 shows an example with several recollision events occurring during one optical cycle.

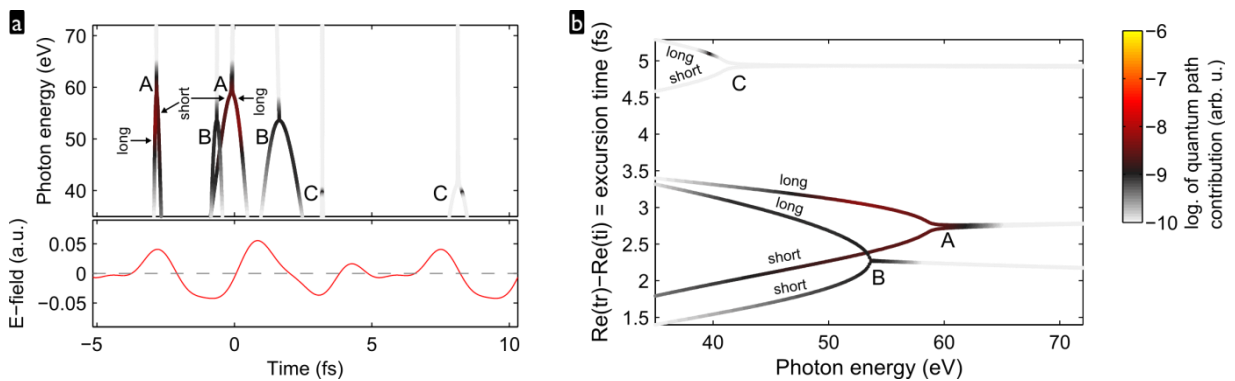

**Figure S2**: Quantum paths calculated for the same parameters as in Fig. 3 of the main paper but with $\phi_2=1.19\pi$ and $\phi_{CEP}=0.5\pi$. (a) Real parts of the ionization and recollision instants presented together with the driving waveform in the same way as in Fig. 3a of the main paper. The three recollision events are named alphabetically and the short and long trajectory branches are marked for event 1. (b) Corresponding excursion times, i.e. the difference the recollision and ionization instants.

**Table S1**: Parameters of the driving waveforms considered in Fig. Figure **S3**. All driver waves have the same fluence per 10.3-fs time window: that of a monochromatic wave with 1.2e14 W cm$^{-2}$.

| Color components | R | φ (rad) |
|---|---|---|
| 1.0 µm + 0.5 µm | 5:1 | φ = 0.0 π |
| 1.5 µm + 0.75 µm | 5:1 | φ = 0.0 π |
| 1.5 µm + 0.5 µm | 3:1 | φ = 1.5 π |
| 1.5 µm + 1.0 µm | 1:1 | φ = 1.575 π <br> (ϕ$_{CEP}$ = 0.85 π in the notation of the paper) |
| 1.5 µm + 1.0 µm + 0.5 µm | 1:1:0.05 | φ$_{1.0}$ = 1.575 π,   φ$_{0.5}$ = 1.35 π <br> (ϕ$_{CEP}$ = 0.85 π and ϕ$_2$=0.2 π in the notation of the paper) |

## 4 Improved HHG optimization by more advanced shaping of the driving waveforms

We have performed quantum path Lewenstein-model simulations in order to understand how the optimization/enhancement of HHG can be improved with our 3-color driver waves as compared to optimized 2-color drivers. For each 2-color combination, we varied the intensity-ratio R (from 0.1 to 10) of the two colors and the phase delay φ (from 0 to 2π) of the shorter-wavelength component to find the optimal parameters in each case. For the 3-color field, we used the intensities estimated for our experiments (Fig. 3 of the main paper and Figure S9) as well as the optimal phase delays for highest efficiency in the cutoff region (same as in Fig.1c of the main paper). The identified optimal parameters (i.e. high cutoff, high efficiency, effective temporal gating) are collected in Table S1. All driver waves have the same fluence per 10.3-fs time window: that of a monochromatic wave with 1.2x10$^{14}$ W cm$^{-2}$. Only the short trajectory class was included in the HHG spectra.

Figure S3 presents the resulting HHG spectra together with the corresponding attosecond pulses after selection of a 20 eV wide slice of the spectrum. This shows the timing and intensity of the attosecond emission within the considered 10.3 fs time window.

The reference is our optimal 3-color waveform (relative fluences as estimated for our experiments in Fig.4 of the main paper, phase delays optimized for HHG photon energies <100 eV), which was already compared to single-color drivers in Fig. 1 of the main paper and is shown again here in Figure S3i,j). It leads to a conversion efficiency similar to that of a monochromatic 1-µm driver, but with a strongly enhanced spectral extension (up to 130 eV) and concentrates the generated XUV energy in a single attosecond pulse per 10.3-fs time window.

As shown in Figure S3a,b), the ω+2ω combination from 1 µm + 0.5 µm in a configuration that enables the "double optical gating" method[7], i.e. it leads to emission of a single attosecond pulse per driver cycle, does lead to a much higher conversion efficiency – but the attainable spectral extension is limited to <60 eV. Using these lower photon energies, the highest attosecond pulse intensities in our comparison is reached, but there are three attosecond pulses emitted in the considered 10.3-fs time window.

Transferring this driver to longer wavelengths, namely to a 1.5 µm + 0.75 µm combination (Figure S3c,d)), improves both on the spectral extension (<95 eV) and the temporal gating (2 attosecond pulses per 10.3-fs time window), but at the expense of conversion efficiency and attosecond pulse intensity. Also, in terms of total attainable pulse energy, this driver would of course be much more costly to be realized in the lab, since it is limited by the total energy of the OPA signal wave.

Considering an ω+3ω combination from 1 µm + 0.5 µm (Figure S3(e,f)) (similar to what has been experimentally realized in ref. 8), we find a very high conversion efficiency again, even with slightly improved spectral extension, but less interesting temporal gating (four attosecond pulses per 10.3-fs time window). Generating this waveform would also not be any easier than the scheme we present in this paper. The considered optimal intensity ratio of the two colors demands exceptionally high conversion efficiency to the third harmonic. Generating it directly from the OPA signal wave would be hard and in any case inefficient in terms of total attainable pulse energy. Generating the 0.5-µm component as the second harmonic of the 1-µm pump beam, as we do in our scheme, is much more efficient, but of course just as technically demanding as our three-color scheme: CEP-locking of the pump beam is necessary for a stable shaped waveform.

Finally, a 2-color combination from 1.5 µm + 1.0 µm (Figure S3g,h)) already unites many of the benefits of our optimal 3-color driver: our simulations result in almost the same spectral extension (<130 eV), temporal gating to a single attosecond pulse per 10.3-fs time window, and a conversion efficiency improved by about one order of magnitude as compared to a monochromatic 1.5-µm driver (not shown). Again, realizing this waveform requires the same effort as our three-color driver, short of the very simple second harmonic generation step. Comparing to Figure S3i,j), we find that by adding the third color, 0.5 µm, with the right phase delay yields another factor 4 in terms of conversion efficiency and attosecond pulse intensity, with very little added experimental complexity.

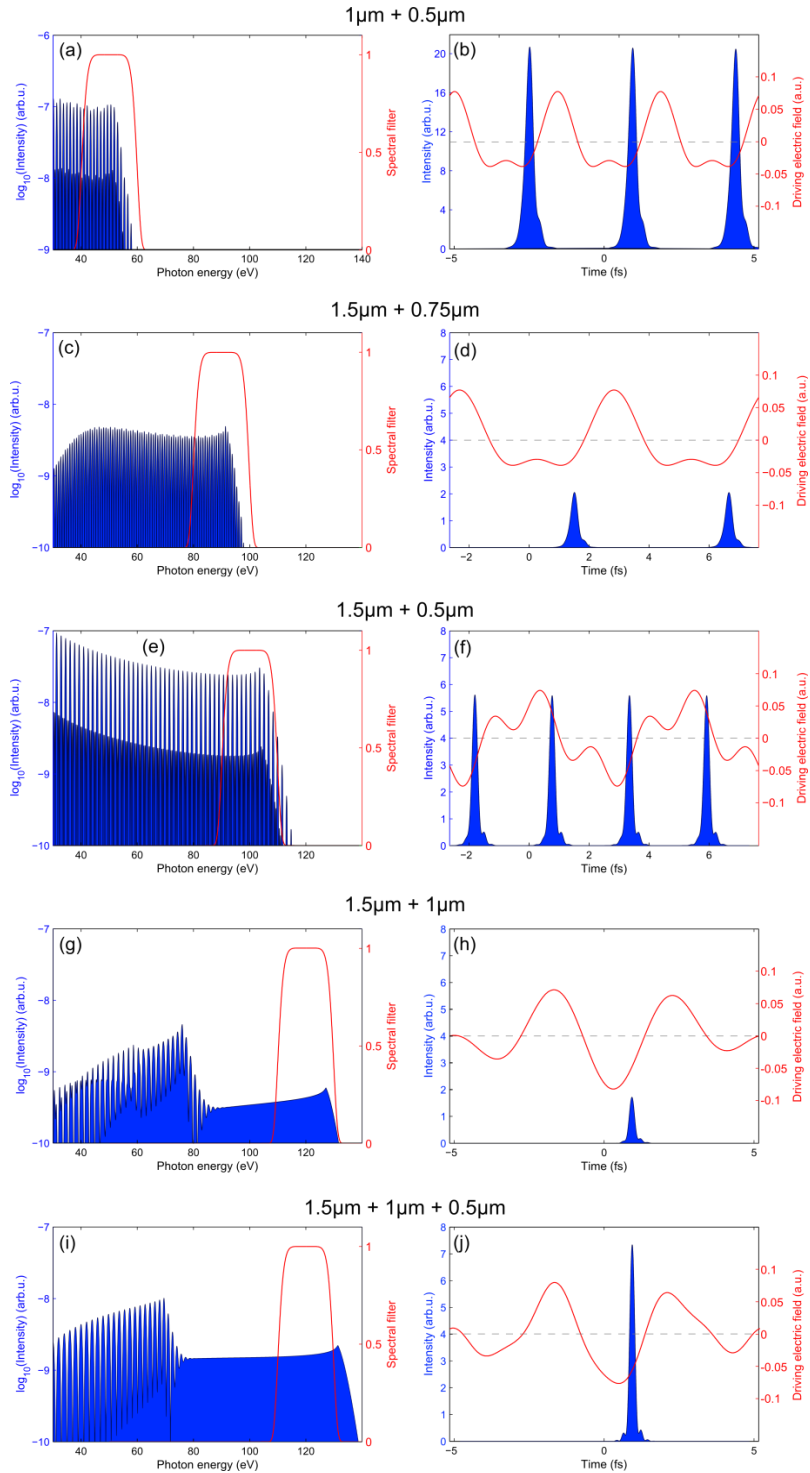

**Figure S3**: Calculated HHG spectra with a considered 20 eV wide filter function (left) and XUV attosecond pulses (right) corresponding to the filtered spectra for the five driving waveforms with parameters as given in Table S1. The two-color drivers are composed from: 1 µm + 0.5 µm (a,b), 1.5 µm + 0.75 µm (c,d), 1.5 µm + 0.5 µm (e,f), 1.5 µm + 1.0 µm (g,h), and the optimal three color wave from 1.5 µm + 1 µm + 0.5 µm (i,j),

In conclusion, the only driver waveforms that would be significantly easier to realize than our 3-color-driver are the ω+2ω combinations because CEP-locking is no longer required (unless of course one wants to apply the double optical gating method [8]). However, using the 1030-m laser for the ω-component severely limits the attainable XUV spectral extension. Using a longer wavelength generated by OPA instead will make it very difficult to reach the same pulse energy level as we do with our 3-color waveform, and even then the 2-color driver remains inferior in terms of XUV spectral extension, attosecond pulse intensity and temporal gating.

The improvement we observe in the simulation can be understood fairly well with very qualitative arguments. By going from an ω+2ω combination to 2ω/3+ω, and further to our 2ω/3+ω+2ω, we gradually approach better the "perfect wave" for HHG[8], theoretically predicted by some of us in 2009. The ω+2ω wave already presents a strong and somewhat "sharpened" ionizing field crest launching recolliding electron trajectories, and the following accelerating field crest is broad as suggested by the "perfect wave" ideal. The accelerating field is however rather weak. This causes the limited extension of the XUV spectra and, due to the imbalance between the strong field first tearing the electron away from the core and that stopping it and accelerating it back, the trajectory excursion durations are actually longer than those for a single-color ω driver. Also, it is still the case that all trajectories launched before the ionizing field peak do *not* recollide. Using the incommensurate 2ω/3+ω combination creates a better balance between ionizing and accelerating field crests, leading to reduced trajectory excursion durations, and the observed enhancement is essentially reached by concentrating the field fluence of a 10.3 fs long time window into two strong field crests. The waveform is however still essentially that of a few-cycle pulse with sinusoidal carrier, as is also apparent from the identity cos(x) + cos(y) = 2 cos[(x+y)/2] cos[(x-y)/2]. Adding the third color component, 2ω, now alters the waveform shape to approach closer the "perfect wave": the ionizing field crests becomes "sharper" and is further enhanced, while the accelerating field crest is made broader and obtains more of the optimal "ramped-up" shape of the "perfect wave". It now also becomes possible to make trajectories launched before the ionizing field peak recollide, thus using most effectively the peak of the strong-field-ionization rate.

How close does this take us to the actual "perfect wave", which is a sawtooth wave requiring many more high-frequency Fourier components to be synthesized? In ref. 8, a realistic approximation to this ideal composed of five colour components, 0.5ω+ω+2ω+3ω+4ω, was proposed. This can be considered as the benchmark for the HHG driving performance of any realistic waveform. Figure S4 compares calculated HHG spectra of this benchmark with a single-colour 1545 nm driver (the low-end benchmark) and two 3-colour drivers with the colour components used in this work. We applied the same genetic algorithm as in ref. 8 to find the optimal combination (relative fluencies and phase delays) of our available three colours. This lead to an optimal wave with 64% of the total fluence in the 1.5-µm component, 28% in the 1.0-µm component, 8% in the 0.5-µm component, and $\phi_{CEP}$ = 0.85π and $\phi_2$ = 0.12π). It leads to a spectral cutoff only slightly lower than that of the benchmark and to a spectral intensity lower by a factor < 2. Experimentally, we have to make a compromise between a high relative weight of the 1.5-µm component and a high total pulse energy (the energy in the 1.5-µm pulse is limited). We therefore raised the relative fluence of the 1.5 µm component only to 50%. Also, the relative fluence of the 0.5 µm component was only ≈ 2.5% in our experiments. The optimal wave with these relative fluences is the one we have shown already in Fig.1 of the main paper and in Fig. S3i,j) and it performs almost as well as the found optimum. At equal total fluence, this waveform, which is within the parameter space of our current experimental setup, thus leads to an only about 10 eV lower spectral cutoff and to spectral intensities reduced by a factor < 3 as compared to the "perfect wave" benchmark.

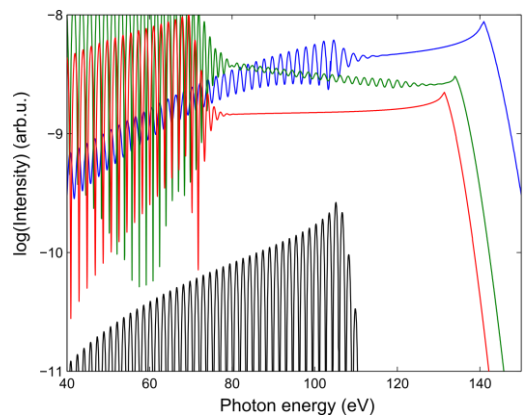

**Figure S4**: HHG spectra calculated over a single period T of the respective driving waveforms and then scaled by ρ=(T$_0$/T)², with T$_0$ = 10.3 fs. The result may be considered as the envelopes of the HHG spectra generated by a *N* periods, where *N* is the least common multiple of the periods of the waveforms to be compared. *Black line*: for a 1545-nm single-colour driver (ρ=4) with a 1.2x10$^{14}$ W cm$^{-2}$ peak intensity. With the same total fluence per 10.3-fs period and a base-frequency ω corresponding to 1030-nm wavelength: *Blue line*: for the 0.5ω+ω+2ω+3ω+4ω combination (ρ=2.25) shown to be a very good approximation to the "perfect wave" in ref. 8 (blue line). *Green line*: for the optimal (2/3)ω+ω+2ω combination (ρ=1) found by the same genetic algorithm as applied in ref. 8 (same as in Fig. 1 of the main paper). *Red line*: for the (2/3)ω+ω+2ω combination (ρ=1) with the relative fluences as estimated for our experiments (Fig. 4 of the main paper) and the phase delays corresponding to φ$_{CEP}$ = 0.85 π and φ$_2$=0.2 π (same as already shown in Figure S3i,j).

## 5 Data analysis

The experimental continuum electron wave packet (REWP) spectrograms shown in Fig. 2 of the main paper are obtained from the measured HHG spectra as described in Figure S5. The recombination dipole matrix element we use to correct for the influence of the atomic structure of argon was calculated by B. Fabre for Ref. 9. The characteristic Cooper minimum around 53 eV photon energy clearly appears in our data. The slowly varying spectral envelope of the spectra, which is the quantity plotted in Fig. 2 of the main paper, is found by performing a peak search for the harmonic peaks and subsequent cubic spline interpolation. This traces the shape of the spectra more accurately than simple smoothing or Fourier filtering.

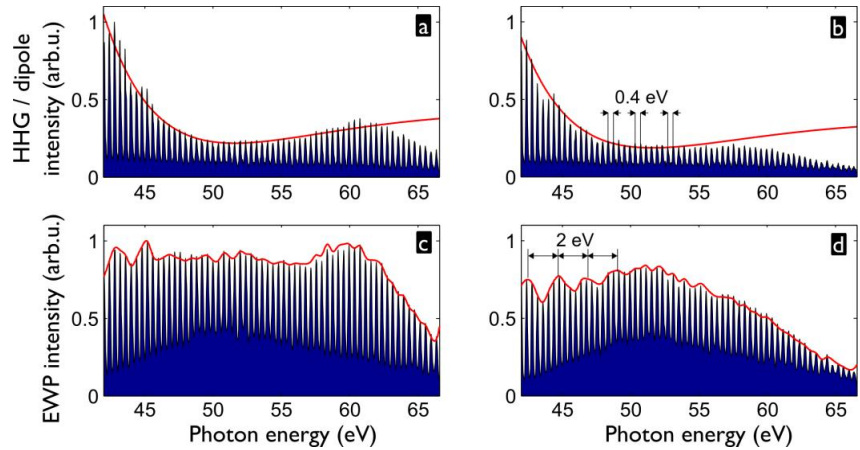

**Figure S5**: (a) and (b): Measured HHG spectra (blue area), taken out of the phase delay-scan shown in Fig. 2 in the main paper, for $\phi_2$ = 1.19π and for $\phi_{CEP}$ =1.6π and $\phi_{CEP}$ = 2.6π, respectively. The red line shows the squared recombination dipole matrix element for argon. (c) and (d):The same HHG spectra divided by the squared recombination dipole matrix element for argon give the corresponding EWP spectra. The red line shows the slowly varying spectral envelope.

## 6 More detailed comparison of experimental and simulated REWP

Figure S6 and Figure **S7** show a more detailed comparison of measured and simulated REWP for a few selected waveforms out of Fig.2 of the main paper. When making this comparison, one should however keep in mind three issues: *(i)* The experimental CEP-values are not known precisely since the CEP-ramp applied during the recording of the spectrograms for each $\phi_2$ –value are not perfectly linear due to non-compensated jitter on the ~1 Hz-level (see Figure S8). So while the unknown offset of $\phi_{CEP}$ may be approximately determined by matching the qualitative spectrogram patterns, there is still a "jitter" on the CEP-step size. *(ii)* The Lewenstein-model, which implicitly includes a static tunneling description, can usually not be expected to yield accurate spectral envelope-shapes—especially not on a linear scale. In a "non-adiabatic tunneling"-conditions (say, Keldysh parameter 0.5 < γ < 2) that all HHG-experiments actually work in, the instantaneous ionization rate depends less steeply on the laser field strength (see Fig.1 in ref. 10). Consequently, the Lewenstein-model predicts an exagerrated modulation-depth of the HHG spectral intensity with the CEP (which varies the strength of the ionizing field crest). *(iii)* More importantly, even when including a certain CEP-jitter (as done in Figure S7), our single-atom calculations over a single laser period neglect a number of effects that all "average out" the deep modulations of HHG. The spatio-temporally varying macroscopic phase matching the ionizing HHG medium does of course shape the HHG spectral envelopes. Already on the single atom-level, non-parallel temporal phases of the colour components (which are not perfectly Fourier-transform-limited), their not-perfectly-parallel wavefronts, as well as the ionization-induced dispersion in the HHG medium all induce additional averaging over $\phi_{CEP}$ and $\phi_2$ (see SM2). Furthermore, the field strength in successive laser cycles obviously varies under the femtosecond envelope, which effectively smoothes the HHG spectral envelopes along the photon-energy axis (see Figure S6d)).

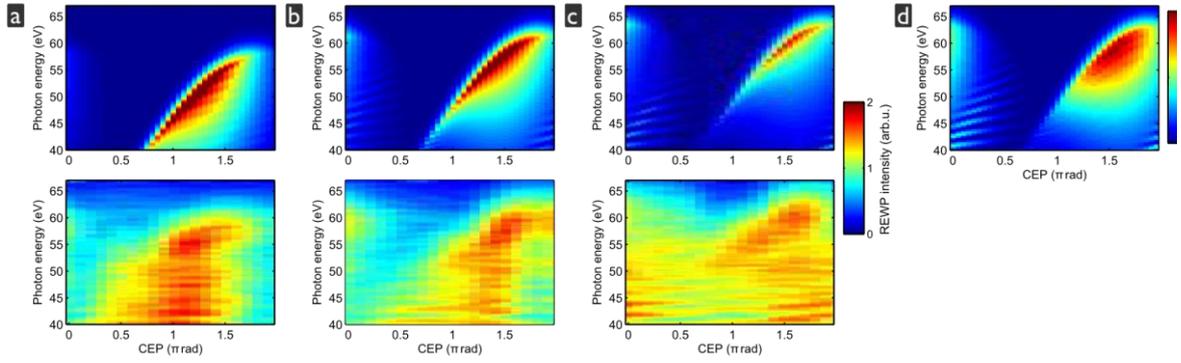

**Figure S6**: Simulated (upper) and measured (lower) REWP spectrograms spectrograms for $\phi_2$ = 0.8π (a), 1.03π (b) and 1.19π (c). These data are zoomed 2π–wide slices out of the spectrograms shown in Fig. 2 of the main paper, shown here for a more detailed comparison. Panel (d) shows a simulation for the same parameters as (c), but for a full pulse with 180 fs FWHM duration instead of just one period.

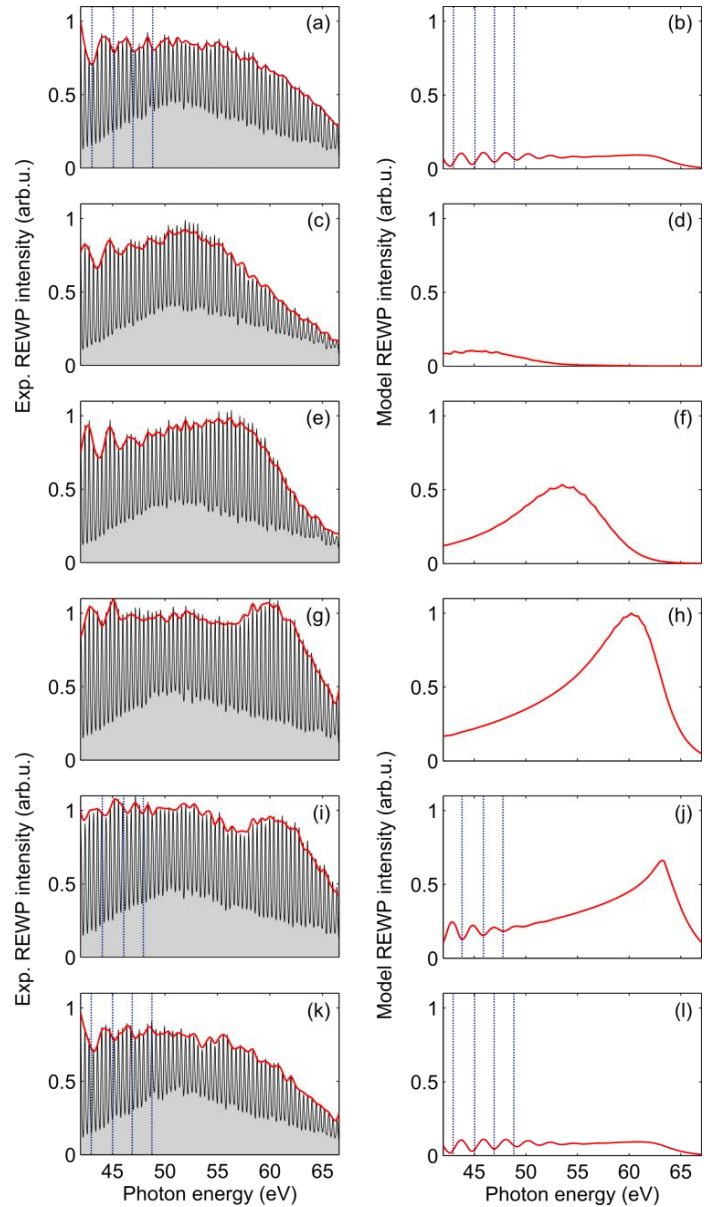

**Figure S7**: Comparison of REWP spectra for the conditions of Fig. 2 of the main paper with $\phi_2$ = 1.19π. Measured (left column) and simulated (right column) spectra are shown for $\phi_{CEP}$=0.4π (a,b), 0.8π (c,d), 1.2π (e,f), 1.6π (g,h), 2.0π (i,j), and 2.4π (k,l). The simulated spectra include a CEP-jitter of 1 rad FWHM. The dotted lines are at the same position in the corresponding measured and simulated spectra and guide the comparison of the slower spectral modulation (≈ 2 eV period) at lower photon energies.

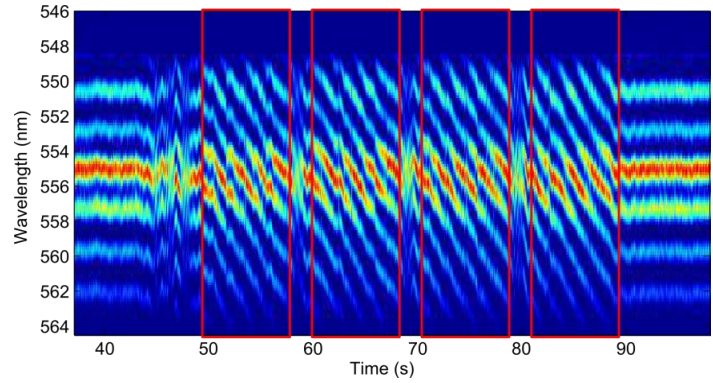

**Figure S8:** Fringes of the f-2f interferometer of the CEP-locking slow-loop (see Fig. Figure **S1**), recorded during a series of four CEP-scans (marked by red frames), each over 8π, like those in Fig. 2 of the main paper.

In conclusion, whilst detailed quantitative agreement of our simulations with the measured HHG spectra cannot be expected, very good qualitative agreement is found. We could include many of the above mentioned issues in more involved simulations by, e.g., including some phase-delay jitter or making the calculations for the whole 180-fs laser pulse instead of a single period. However, the point we want to make with our simpler simulations and the agreement we do indeed obtain in Fig. 2 of the main paper is that the qualitative structures appearing in the measured spectrograms (photon-energy vs. $\phi_{CEP}$) have their origin in single-atom sub-cycle dynamics. This confirms that we do exert control on this level via our shaped waveforms. For any quantitative question, such as "How much enhancement will we get in the laboratory in a given spectral region?", we may get rough expectations from our model calculations, but the real answer requires experiments such as those reported in this paper.

## 7 Phase delay scans at higher intensity

Figure S9 shows in analogy to Fig. 2 of the main paper the phase delay scans under the higher-intensity conditions. The HHG spectrum shown in Fig. 4c of the main paper has been selected from these scans.

While there is some agreement between the overall structures of the simulated and measured spectrograms for the different $\phi_2$-values, the agreement is much reduced compared to that for the lower intensity case discussed earlier. From this we conclude that macroscopic effects in the HHG medium play a major role in the shaping of these spectra. Higher photon energies seem to be less well phase matched. An analysis of these complicated and often highly dynamical macroscopic effects goes beyond the scope of this paper. The purpose of this part of the study is to show that even under experimental conditions where the single atom response can no longer alone explain the features of the macroscopic HHG signal, our shaped optical cycles are an effective means to enhance HHG.

Also, the qualitative matching of the simulated and measured spectrograms again allows us to determine approximate values for the experimentally unknown offsets of the relative phases $\phi_2$ and $\phi_{CEP}$. Based on this, we estimate the values $\phi_2=1.8\pi$ and $\phi_{CEP}=0.3\pi$ for the waveform that generated the enhanced HHG spectrum in Fig. 4c of the main paper.

For this waveform, Figure S10 shows the calculated quantum paths, as well as the attosecond pulse intensity corresponding to a 20-eV-wide slice around 60 eV photon energy from the spectrum. A single event, with ionization at the strongest field crest, by far dominates the HHG emission. Therefore, essentially all of the measured integral XUV flux in Fig. 4c of the main paper corresponds to a single attosecond XUV pulse per 10.3-fs period.

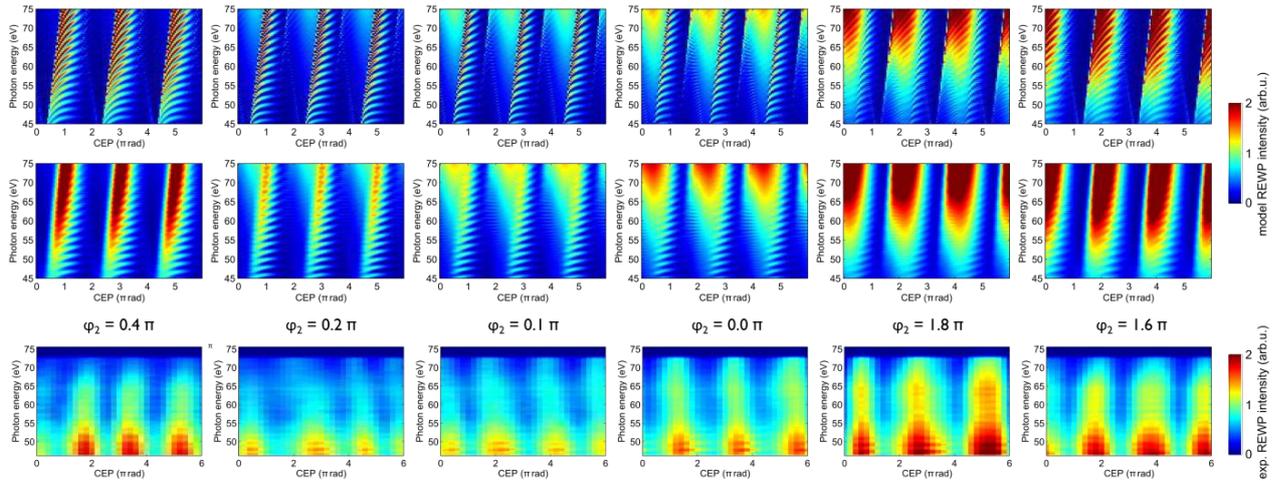

**Figure S9**: Simulated (two upper rows) and measured (lower row) EWP spectrograms for the $\phi_2$-values given for the different columns. In the simulations, only the short trajectory family has been included. The difference between the upper and the middle row is that in the latter, a 1 rad wide Gaussian averaging over the CEP has been applied to simulate CEP jitter/ propagation-induced averaging.

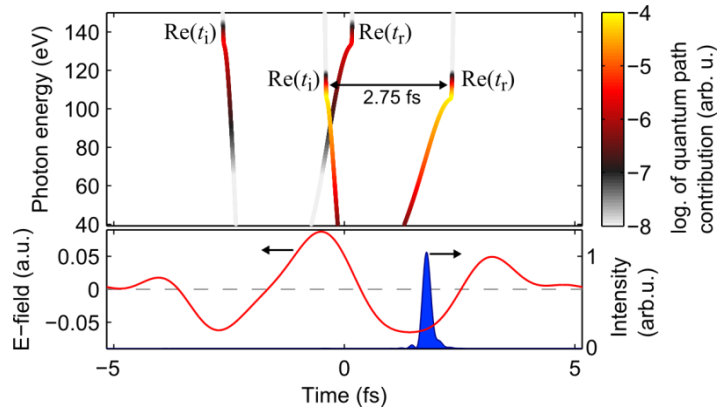

**Figure S10**: Results of the saddle point analysis of the Lewenstein-model simulations for the parameters of Figure S9 with $\phi_2=1.8\pi$ and $\phi_{CEP}=0.3\pi$, which also represents the experimental 3-colour driver chosen for Fig. 4c of the main paper. The upper panel shows ionization and recollision instants for the two recollision events occurring per 10.3-fs period. Only the short trajectories are shown. The lower panel shows the driver waveform as well as the intensity envelope of the attosecond pulse resulting from selecting a 20-eV-wide spectral slice around 60 eV photon energy from the calculated spectrum, which results from coherent addition of both recollision events. The time axes of upper and lower panel are the same.

## 8 Saturated HHG with 1030-nm driver

Figure S11 shows a series of HHG spectra generated in argon with the sinusoidal 1030-nm driver with different pulse energies. Increasing pulse energies lead to the increasing ionization of the HHG medium, which leads to a range of (often highly dynamical) effects limiting the achievable spectral extension and efficiency of HHG. Such effects include: saturation of the effective HHG driving intensity due to plasma defocusing, depletion of the HHG medium ground state before the pulse intensity peak, and the deterioration of phase matching due to plasma dispersion. From 0.54 mJ pulse energy on, which turns out to be the same value as the total energy of the synthesized pulses used for Figure S9 and Fig. 4c of the main paper, the observed HHG cutoff remains below the value that would be expected if the effective driving intensity increased linearly with the pulse energy. We have increased the driving pulse energy well beyond this value, but saturation effects limit the attainable HHG cutoff to ≈ 60 eV. These saturation effects depend on the pulse duration and shorter pulses would allow higher saturation intensities.

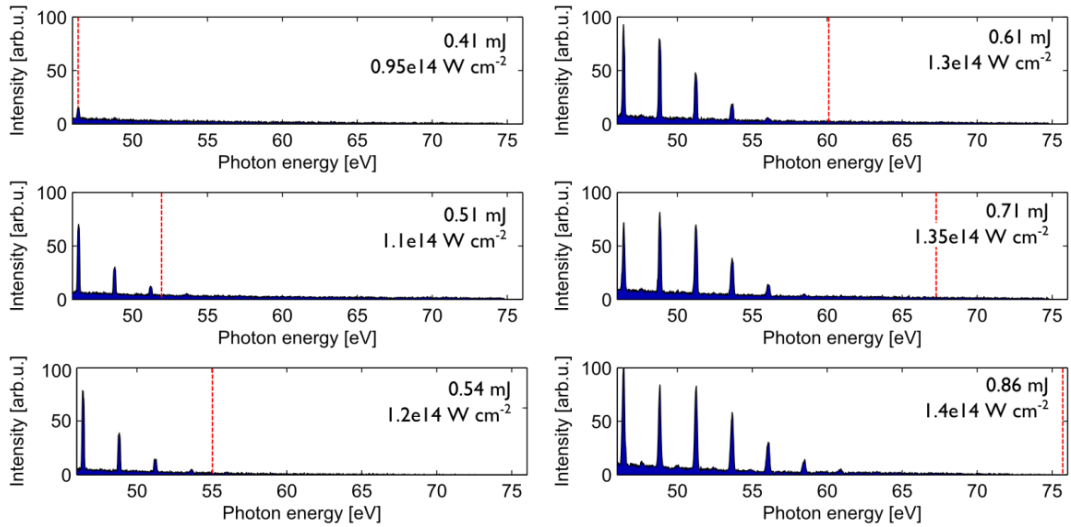

**Figure S11**: HHG spectra generated in argon with the 1030-nm driver in the same conditions as those in Fig.4 of the main paper. Only the pulse energy, given in each panel, has been varied. The dashed red lines mark the HHG cutoff position as it should increase in the absence of saturation effects, i.e. if the effective driving intensity increased linearly with the pulse energy. Also given in each panel is the apparent effective driving intensity extracted from the observed cutoff positions.